\documentclass[twocolumn]{aa}
\usepackage{graphicx}
\begin{document}
   \title{Constraining the properties of spots on \\Pleiades very low mass stars}

   \author{Alexander Scholz$^{1,3}$\thanks{Visiting Astronomer at the German-Spanish
   Astronomical Centre, Calar Alto, operated by the Max-Planck-Institut f\"ur
   Astronomie, Heidelberg, jointly with the Spanish National Commission for
   Astronomy.}
   \and
   Jochen Eisl{\"o}ffel$^1$
   \and
   Dirk Froebrich$^2$}

   \offprints{A. Scholz, e-mail: aleks@astro.utoronto.ca}

   \institute{$^1$ Th{\"u}ringer Landessternwarte Tautenburg,
              Sternwarte 5, D-07778 Tautenburg, Germany\\
              $^2$ Dublin Institute for Advanced Studies, 5 Merrion Square, Dublin
              2, Ireland\\
	      $^3$ University of Toronto, Department of Astronomy \& Astrophysics, 60 St. George Street, 
	      Toronto, M5S 3H8, Canada}
	      
   \date{Received sooner; accepted later}
   
   \authorrunning{A. Scholz et al.}
   
   \titlerunning{Spots on very low mass stars}

   \abstract{We present results of a multi-filter monitoring campaign for
   very low mass (VLM) stars in the Pleiades. Simultaneous to our I-band
   time series (Scholz \& Eisl{\"o}ffel \cite{se04b}), which delivered photometric
   periods for nine VLM stars, we obtained light curves in the J- and H-band. 
   One VLM star with $M\approx0.15\,M_{\odot}$ (BPL129) shows a period
   in all three wavelength bands. The amplitudes in I, J, and H are 0.035, 0.035, 
   and 0.032\,mag. These values are compared to simulations, in which we
   compute the photometric amplitude as a function of spot temperature and
   filling factor. The best agreement between observations and models is found
   for cool spots with a temperature contrast of 18-31\% and a very low surface filling
   factor of 4-5\%. We suggest that compared to more massive stars VLM objects may 
   have either very few spots or a rather symmetric spot distribution. This difference 
   might be explained with a change from a shell to a distributed dynamo in the
   VLM regime.
   }

   \maketitle

\section{Introduction}
\label{intro}

Magnetic activity is an ubiquitous phenomenon on late-type stars. It shows up as
coronal X-ray emission (e.g., Feigelson \& Decampli \cite{fd81}, Schrijver et al.
\cite{smw84}, Stauffer et al. \cite{scg94}) or as chromospheric emission in lines 
like H$\alpha$ and CaH\&K (e.g., Wilson \cite{w78}, Soderblom \cite{s85}, Henry et al. 
\cite{hsd96}). In the photosphere, the most obvious indication for magnetic activity 
is the existence of dark star spots. 

Probably the best way to investigate properties and evolution of these magnetic spots
is to construct two-dimensional Doppler images from a spectroscopic time series. Such 
Doppler images are available for a number of late-type stars at various evolutionary
stages, and they clearly show the existence of large, cool spots, which are in most cases
asymmetrically distributed (e.g., Vogt et al. \cite{vph87}, Piskunov et al. \cite{ptv90}). 
A more indirect way to investigate surface spots is photometric monitoring: Late-type 
stars often show a periodic light curve, and the best explanation for this behaviour is 
the flux modulation by cool spots co-rotating with the objects (e.g., Bouvier \& Bertout
\cite{bb89}, Strassmeier \cite{s92}). It is possible to assess the spot properties, 
such as temperature and size, by monitoring large samples of targets in more than one filter 
as done by, e.g., Herbst et al. \cite{hhg94} and Bouvier et al. \cite{bck95} in the case
of T Tauri stars.

In the last few years, it has become clear that phenomena which
are usually attributed to magnetic activity do also occur on very low mass
(VLM) stars, i.e. stars with masses below $0.4\,M_{\odot}$. At least down
to spectral types of M7-M9, these objects show both X-ray (Mokler \& Stelzer 
\cite{ms02}) and H$\alpha$ emission (e.g., Mohanty \& Basri \cite{mb03}), indicating 
the presence of coronal and chromospheric activity. A significant fraction of
VLM objects also shows periodic photometric variability, and it is
commonly believed that these variations have their origin in magnetic
surface spots (e.g., Herbst et al. \cite{hbm02}, Scholz \& Eisl{\"o}ffel 
\cite{se04a}, Lamm et al. \cite{lmb05}).

In the past, little information has been obtained about the properties of spots on
VLM stars. This is mainly due to the fact that these objects are very faint, which 
makes it difficult to obtain high signal-to-noise spectra and light curves. 
Such studies are, however, very interesting, since for
two reasons we expect a change of the spot characteristics in the VLM regime: a) 
Stars with $M<0.35\,M_{\odot}$ are believed to be fully convective (Chabrier \& 
Baraffe \cite{cb97}). Therefore, they cannot form a solar-type $\alpha\Omega$-dynamo, 
which operates in a shell between convective and radiative zone (Parker \cite{p75}, 
Spiegel \& Weiss \cite{sw80}). Thus, there has to be a different mechanism for 
magnetic field generation, which may have consequences for the spots. b) VLM objects 
have very low effective temperatures compared to solar-mass stars. This reduces 
the coupling between magnetic field and gas. Thus, we should expect a decrease 
of magnetic activity as the objects become cooler. 

Only recently, the advent of wide-field detectors made it possible
to obtain large samples of periodic light curves of VLM objects, mainly for targets 
in very young clusters like the Trapezium cluster (Herbst et al. \cite{hbm01}, 
\cite{hbm02}) and \object{NGC2264} (Lamm \cite{l03}, Lamm et al. \cite{lbm04}, 
\cite{lmb05}). From our own monitoring campaigns, we published about 60 new 
periods for VLM objects in the clusters \object{$\sigma$ Ori} (Scholz \& 
Eisl{\"o}ffel \cite{se04a}), \object{Pleiades} (Scholz \& Eisl{\"o}ffel \cite{se04b},
hereafter SE), and \object{$\epsilon$ Ori} (Scholz \& Eisl{\"o}ffel \cite{se05}). 
These studies give evidence for a change of the spot properties at very low masses: 
In the young cluster NGC2264, the amplitudes in the I-band lightcurves of the 
(non-accreting) VLM stars are on average reduced by a factor of 2.7 in comparison 
with more massive stars (Lamm \cite{l03}). VLM stars in the Pleiades show a similar 
behaviour. Their average amplitudes in their I-band lightcurves are reduced by a 
factor of four in comparison with V-band amplitudes of more massive stars. Such a 
comparison overestimates the difference in the amplitudes, since for any fixed 
combination of spot temperature and filling factor the amplitudes are always 
smaller in the I-band than in the V-band. However, even after converting 
the I-band amplitudes for VLM objects to the V-band using synthetic spectra, the 
VLM lightcurve amplitudes in the Pleiades are still at least a factor of 2.4 smaller 
than those of solar-mass objects (see Scholz \cite{s04}). Thus, for objects with 
masses below 0.3-$0.4\,M_{\odot}$, corresponding to spectral type M3-M4, the average 
light curve amplitudes tend to be smaller with respect to the higher mass objects 
(although there are still a few M dwarfs with relatively high amplitudes, e.g. 
GT Peg or YZ CMi, see Messina et al. \cite{mrg01}).

Since the distribution of the amplitudes depends only on the spot properties, the 
random orientation of the rotational axes, and the rotational properties, this 
decrease of the amplitudes requires different spot characteristics in comparison with
solar-mass stars in the same rotational regime. At least four scenarios are possible: 
a) low spot coverage, b) low temperature contrast between spots and photospheric 
environment, c) a small degree of asymmetry in the spot distribution, d) most spots 
are polar spots.

Doppler images have been obtained for a few stars with spectral types around M2,
(Barnes \& Collier Cameron \cite{bc01}). These maps show 
many spots distributed mainly at low latitudes, which might exclude the scenarios 
a) and d) above. The stars for which Doppler images have been taken, however,
have masses around or above $0.4\,M_{\odot}$. Thus, they are probably at the upper 
limit of the mass range where the photometric amplitudes decrease. 

The first multi-filter monitoring study for VLM objects has been published by 
Terndrup et al. (\cite{tkp99}). For HHJ-409, a Pleiades star with $M=0.39\,M_{\odot}$,
they measured amplitudes of 0.035 in the I- and 0.08 in the V-band. This is rather
low in comparison with more massive Pleiades stars, though HHJ-409 has a rotation 
period of only 6\,h, i.e. it is in a 'saturated' rotational regime, where actually 
solar-mass stars tend to show decreased amplitudes (Messina et al. \cite{mrg01}). From 
these results, Terndrup et al. (\cite{tkp99}) derive a temperature contrast of 200\,K (6\%) 
and a surface filling factor of 13\% for cool spots on HHJ-409, a Pleiades star with 
$M=0.39\,M_{\odot}$. For comparison, solar-mass stars show surface filling factors 
up to 25\% and temperature contrast between 5 and 40\% (e.g., Strassmeier \cite{s92}). 
The filling factor of the VLM star observed by Terndrup et al. thus is a rather typical 
value, but the temperature difference seems to be quite low. Hence, the result of Terndrup 
et al. (\cite{tkp99}) might indicate that a low temperature contrast is the reason for the 
decreased lightcurve amplitudes on VLM objects. However, this object still has a mass at 
the upper limit of the VLM regime, and it is therefore doubtful whether it is prototypical 
for this object class. Similar observations for objects with masses well below $0.3\,M_{\odot}$ 
are needed.

In this paper, we set out to explore the spot properties for VLM stars in the
Pleiades by analysing their lightcurve amplitudes. Thus, we apply basically
the same technique which has been successfully used for more massive stars
in the past. This paper is structured as follows: Sect. \ref{nirmon} describes 
our photometric monitoring campaign in the near-infrared J- and H-bands, carried 
out in parallel with our published I-band time series in the Pleiades (SE). For 
these three wavelength bands, we determine photometric amplitudes in Sect. \ref{amps}. 
These results are then compared to amplitudes derived from simulations (Sect. 
\ref{models}).

\section{Near-infrared monitoring}
\label{nirmon}

Our I-band time series in the Pleiades was obtained with the CCD camera at the
1.23-m telescope on Calar Alto. It delivered high-precision light curves
for 26 VLM stars in this cluster, from which we were able to determine nine
photometric rotation periods (see SE). To complement these I-band data, 
we aimed for simultaneous monitoring in the near-infrared J- and H-bands to 
determine spot properties. In the following we describe these observations, 
as well as the derivation of the NIR light curves.

\subsection{Observations and image reduction}
\label{obsred}

In the last three nights of our I-band observations (from 17 to 19 October 2002), 
we observed the same target objects also with the 2.2-m telescope on Calar Alto, 
equipped with the near-infrared camera MAGIC (see Herbst et al. \cite{hbb93}). With 
MAGIC in wide-field mode, we obtained a field of view of $7' \times 7'$ and a
pixel-scale of $1\farcs6$/pixel. The images in the I-band have a size of 
$17' \times 17'$, and we observed two fields in the I-band time series. At the 
time of the observations, we did not know which of the targets would be variable, 
thus we tried to cover most of them also in the NIR time series. This was possible 
with eight MAGIC pointings within the two I-band fields. 

\begin{figure}[t]
\resizebox{9.0cm}{!}{\includegraphics[angle=-90,width=6.5cm]{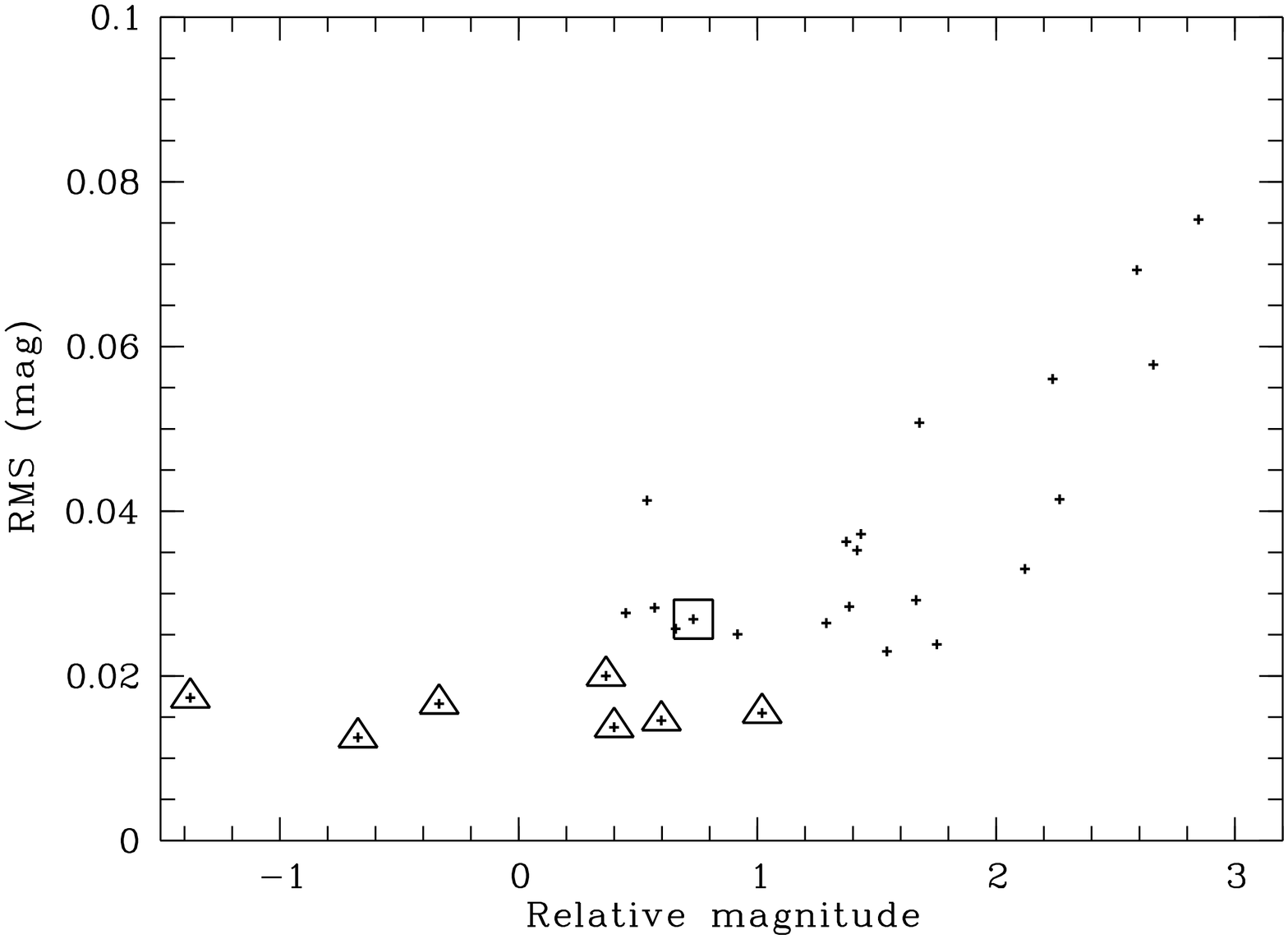}} \hfill
\resizebox{9.0cm}{!}{\includegraphics[angle=-90,width=6.5cm]{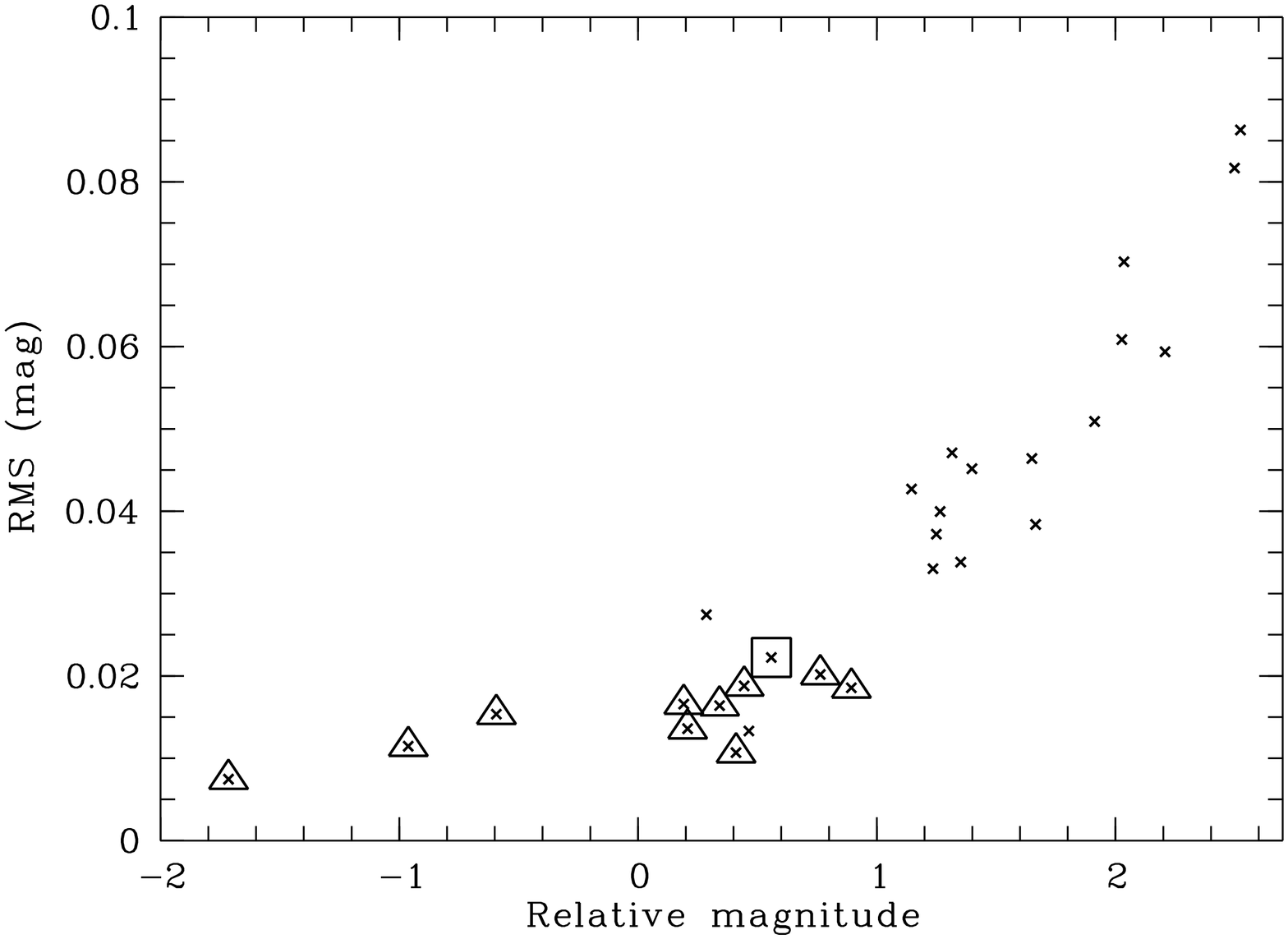}}
\caption{RMS for the J-band (upper panel) and H-band (lower panel) time series.
Small crosses show the values for all sources in the object catalogue. Overplotted
triangles mark the reference stars, which were used for the relative calibration.
The overplotted square marks the object BPL129.}
\label{rms}
\end{figure}

The observations were carried out with the following strategy: We started
with a full J-band sequence over all eight fields, where the exposure time per field
was $13 \times 5$\,s. Subsequently, the same sequence was repeated in the H-band, this
time with exposure times of $19 \times 5$\,s. The first frame of each exposure
was removed because it generally suffers from irregular bias variations. The remaining
frames were summed up, so that we obtain one image per field, filter, and sequence. 
Thus, the exposure times for each data point were 60\,sec in the J- and 90\,sec in the
H-band, which should give us similar signal-to-noise in both filters. 

This whole sequence of 8 fields in J- and H-band, was executed throughout the nights, 
weather permitting. In the first night, observations had to be stopped after five 
and a half sequences because of cloud cover. A software problem restricted the observations 
in the second night to 13 sequences, and in the third night we obtained 14 full sequences. 
Altogether, we have 33 data points for each field in the J-band and 32 in the H-band. Apart 
from the first night, the atmospheric conditions were mostly photometric.

The data reduction of our NIR images uses standard
techniques, but pays attention to the fact that we want to obtain as precise
as possible photometry. 
We bias-corrected all our images and flat-fielded them with skyflats 
constructed by averaging the science frames and rejecting objects with a sigma 
clipping procedure. 
Sky-subtraction was performed using the xdimsum package in IRAF\footnote{IRAF
is distributed by the National Optical Astronomy Observatories, which are
operated by the Association of Universities for Research in Astronomy, Inc.,
under cooperative agreement with the National Science Foundation.} (Stanford et
al. \cite{sed95}). This procedure determines the background for each science frame 
as running average of a number of $n$ images taken before and after the image of 
interest, after excluding the objects.
We performed this procedure for each of our sequences of J- and H-band
images individually. 
The number $n$ of images used in the determination of the sky-background is crucial 
for the final accuracy of our photometry. A larger number of frames 
leads to less noise in the sky background, since more images are averaged. On the 
other hand, the sky background may be variable on timescales on the order of the length 
of our sequence as a consequence of variable weather conditions. Hence, using too many 
images will introduce errors. To obtain the best choice for $n$ we varied this parameter 
from six to eight for each image. For our final photometry the background corrected image 
with the smallest background noise was used. 
For most images the best result was achieved with $n=8$, which confirms that the weather 
conditions were mostly stable at least over one full sequence, which enables reliable
background subtraction for all images of the sequence.

\subsection{Photometry and relative calibration}
\label{photo}

For all eight MAGIC fields, we created an object catalogue from a time series
image with good seeing. Subsequently, we measured instrumental magnitudes by 
aperture photometry. The choice of the aperture radius has to account for the seeing 
during the observations, which varied between 1.0 and 1.8\,pixels in the J-band. In 
the H-band, the average point spread function was about 0.2\,pixels larger, due to a 
slight defocusing. Choosing a fixed aperture of 3\,pixels, we include $>99$\% of the 
stellar flux for all images. 

The relative calibration of the instrumental light curves was carried out following
the recipe presented in Scholz \& Eisl{\"o}ffel (\cite{se04a}). In a first step,
images with unreliable photometry were identified, i.e. images which deliver 
outlying data points for most objects, caused by strong background gradients or 
an excessive amount of cosmics. Note that these images have no influence on
the sky subtraction (Sect. \ref{obsred}), because the running average procedure
in xdimsum excludes 'bad images' reliably. Since they might, however, disturb the 
photometry, we excluded them from further analysis. The basic principle of the 
relative calibration is to construct a mean light curve from a set of non-variable 
reference stars. These stars were selected by comparing the light curve of each 
reference star with the mean light curve of all other 
reference stars. With this procedure, we selected typically 5-10 stars which are non-variable 
with respect to all other stars. Their mean light curve was subtracted from the time series 
of our Pleiades targets. In Fig. \ref{rms}, we illustrate the results of the relative 
calibration by plotting the RMS for all light curves in one MAGIC field. The values for 
the reference stars and a target star (in this case: BPL129) are marked. For the brightest 
objects, we typically reach a mean precision of 15-20\,mmag both in the J- and the H-band.

\section{Amplitude fitting}
\label{amps}

Phased light curves for all Pleiades targets in the I-, J-, and H-band were calculated from
the period and zero point of the I-band time series analysis described in SE.
Since period and zero point are fixed, only amplitudes remain to be determined.
To improve the signal-to-noise ratio, we applied a moving median filter to the
near-infrared phased light curves by averaging over three consecutive data points in
phase space. This procedure is legitimate, because the lightcurve shape does not change
significantly over the observing nights, as can be seen from the high signal-to-noise
I-band time series.

\begin{figure}[t]
\resizebox{9.0cm}{!}{\includegraphics[angle=-90,width=6.5cm]{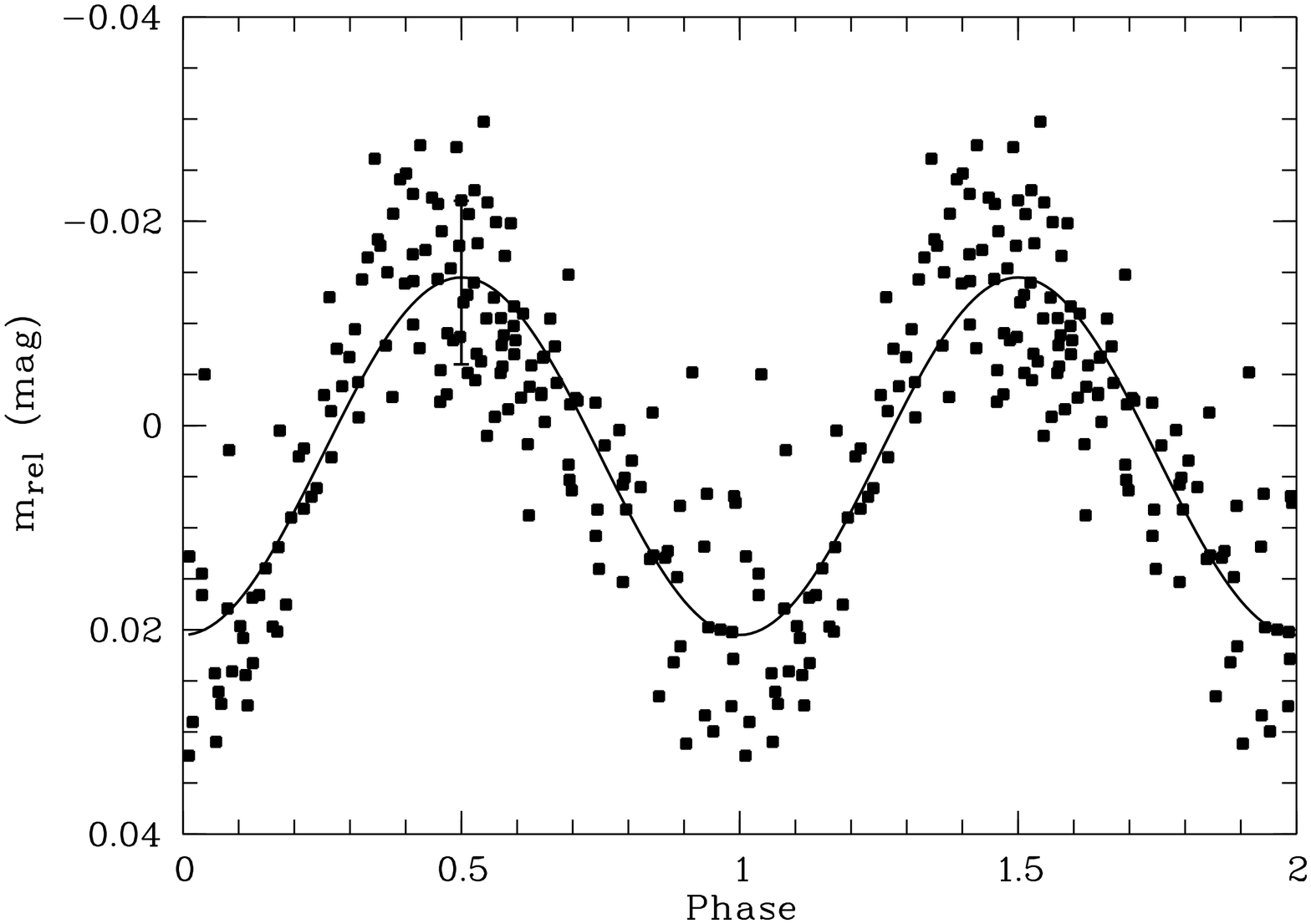}} \hfill
\resizebox{9.0cm}{!}{\includegraphics[angle=-90,width=6.5cm]{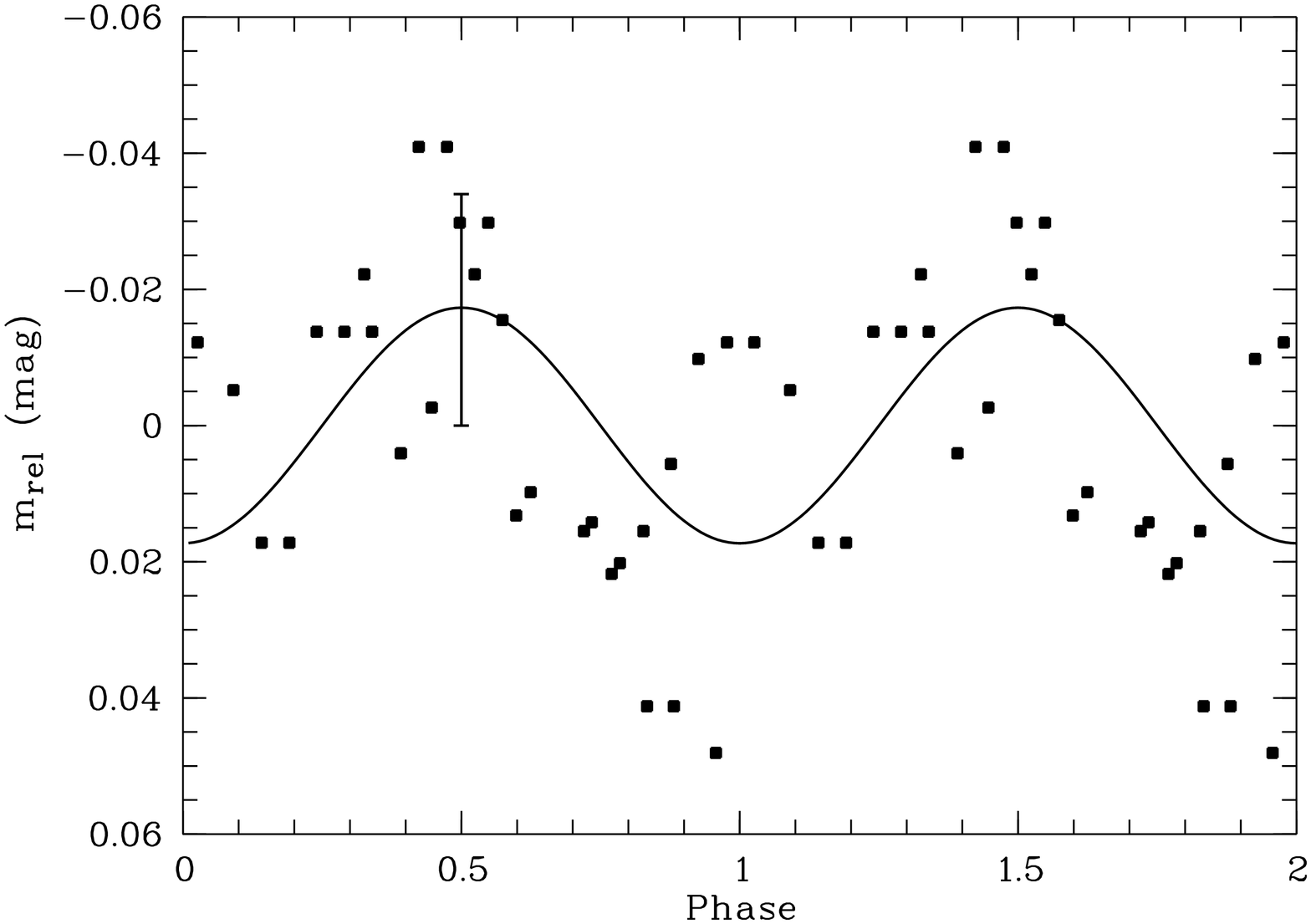}} \hfill
\resizebox{9.0cm}{!}{\includegraphics[angle=-90,width=6.5cm]{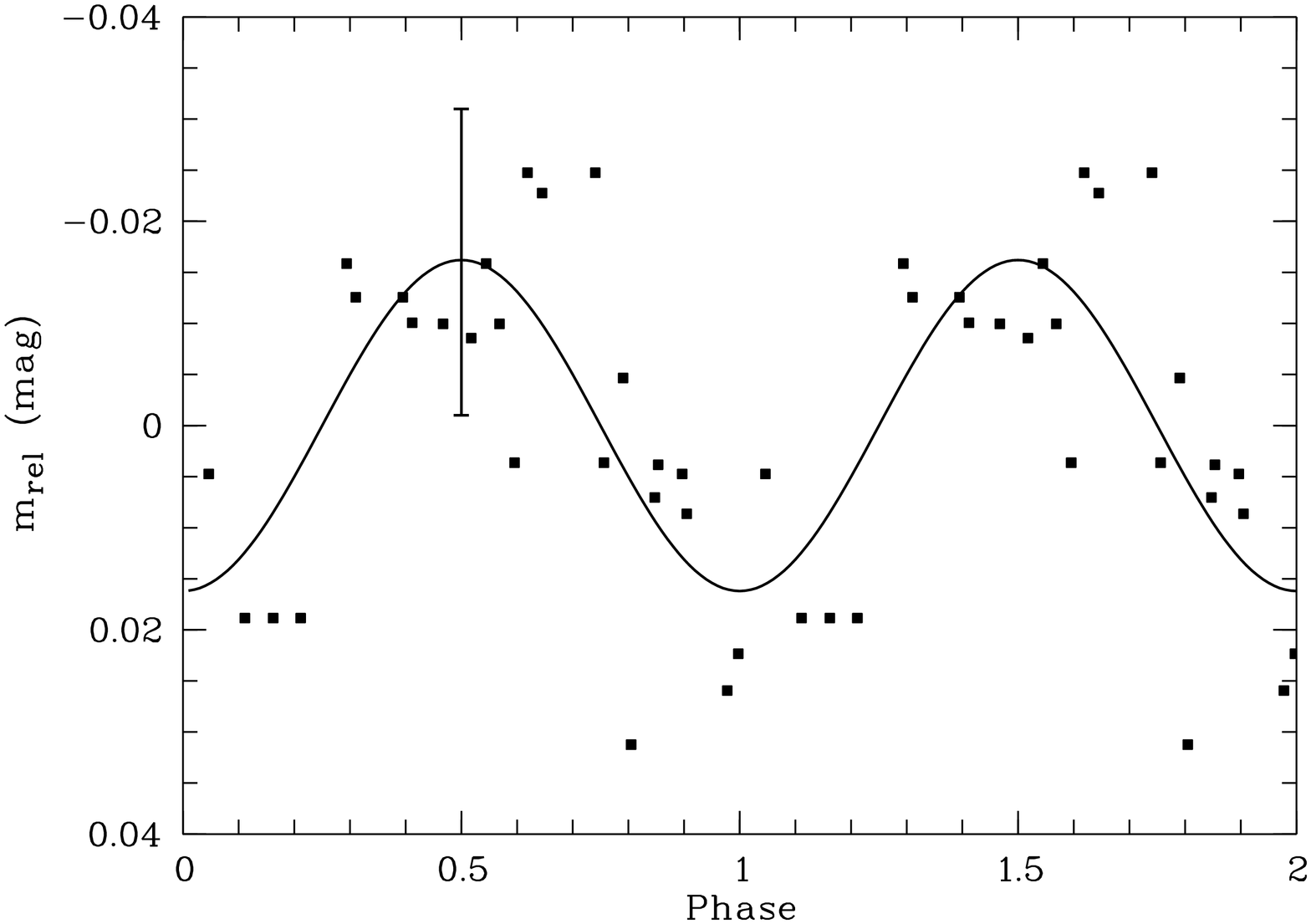}} \hfill
\caption{Phased light curves for BPL129 in the I-, J-, and H-band (from top to bottom).
The solid lines shows the sine waves with the amplitudes for which the residuals are 
minimised. The error bar indicates the uncertainty for the derived amplitude.}
\label{ilc}
\end{figure}

As can be seen in the I-band light curves (SE), the light variations are usually 
well-approximated by a sine shape. Therefore, the amplitudes were determined by 
subtracting sine waves with the given period and zero point from the original 
light curve. We varied the amplitude until the residual noise was minimised. This 
procedure was carried out for amplitudes ranging from 0.001 to 0.1\,mag, for all 
objects and filters.

For eight of our nine targets the signal-to-noise turned out too low
to detect a significant amplitude in the NIR light curves. This is mainly due to the
lower photometric precision in the NIR compared to the I-band, because of lower image
quality and high background. The VLM star \object{BPL129} (catalogue number of Pinfield et al. 
\cite{phj00}), however, shows a clearly defined period in the J- and H-band light curves (see
Fig. \ref{ilc}). In all three filters, the residual noise has a clear minimum. We obtained 
amplitudes of 0.035\,mag for the I-band, 0.035\,mag for the J-band, and 0.032\,mag for the 
H-band. Sine waves with these amplitudes are overplotted in Fig. \ref{ilc}. 

Errors for these amplitudes can be determined from the noise of the residual 
light curve, i.e. after subtraction of the fitted sine wave. Using this method, we 
obtain 0.008\,mag in the I-, 0.020\,mag in the J-, and 0.011\,mag in 
the H-band. This method works well with many data points, when the noise level
is not significantly influenced by low number statistic. Thus, we adopt this
error value for the I-band amplitude, where we have 160 data points. For the 
near-infrared light curves, with less data points, we decided to follow a different 
strategy for their error estimate: Averaging the rms of our non-variable
reference stars (see Sect. \ref{photo}), we are able to obtain a reliable estimate 
for the photometric precision. This gives us values of 0.017\,mag in the J- and 
0.015\,mag in the H-band. Since these values are now based on the light curves of 
several stars, the statistical uncertainties are significantly decreased. We adopt 
these values as uncertainty of our amplitudes in the J- and H-band. We note that all 
these error estimates are conservative, since we do not make use of the fact that the
sine fit uses {\it all} data points of the light curve. Instead, the error
estimate is based on the photometric precision for one single data point.

\section{Surface simulations}
\label{models}

We now compare the observed lightcurve amplitudes with model 
calculations of the surface properties of VLM objects to constrain the 
temperature and the filling factor of spots on these objects. After a description 
of the method in Sect. \ref{desc}, we use the I-band amplitudes for nine
Pleiades VLM stars from SE for a first assessment of the spot properties in Sect. 
\ref{iband}. In the subsequent Sect. \ref{nirband}, we focus on the object BPL129, 
for which we have derived near-infrared amplitudes in Sect. \ref{amps}. Finally, 
we will compare our results with values for more massive stars (Sect. \ref{comp}).

\subsection{Description of the simulation}
\label{desc}

The amplitude of the light curve in this paper is defined as peak-to-peak
magnitude difference. From our observations we have determined this value for
three different wavelength bands. Thus the model should deliver 
the spectral energy distribution (SED) of the target in the maximum and the minimum 
of the light curve. In the following, we define as 'front side' of the star
the hemisphere which lies in the line of sight, whereas the 'backside' is
the hemisphere which is not visible for the observer at a given time. 

The observed SED will be a mixture of the photospheric spectrum and the spectrum 
of the spots, which can either be cooler or hotter than the photosphere. This 
observed SED will change as a function of rotational phase. The prerequisite for 
our simulations is thus to know the SED dependence on effective temperature. 
For this study, we use the model spectra from the Lyon group, namely the models 
'StarDUSTY2000' which are NextGen models (Allard et al. \cite{aha97}), 
improved with new TiO opacities (Allard et al. \cite{ahs00}) and dust opacities 
(Allard et al. \cite{aha01}).

\begin{table}[t]
\caption{Properties of our main target BPL129. Apparent magnitudes are from 
Pinfield et al. (\cite{phj00}) and from 2MASS, absolute magnitudes were derived
adopting $E_{B-V}=0.03$ and a distance of 133\,pc. The masses were obtained by
comparing with the evolutionary track for 125\,Myr of Baraffe et al. (\cite{bca98}).}
\begin{tabular}{lccc} 
\hline
Filter   & m   & M   & Mass ($M_{\odot}$)\\
\hline
I & $16.25\pm 0.02$ & $10.57\pm 0.06$ & 0.14\\ 
J & $14.27\pm0.03$  & $8.61\pm 0.06$ & 0.165\\
H & $13.72\pm 0.05$ & $8.07\pm 0.08$ & 0.16\\
K & $13.40\pm 0.03$ & $7.79\pm 0.06$ & 0.16\\
\hline				     	      
\end{tabular}	
\label{bpl129}		     		     
\end{table}		

We estimate the effective temperature of our main target BPL129 with the available 
photometry: I-band data for BPL129 was published by Pinfield et al. (\cite{phj00}),
and we obtained additional near-infrared data from 2MASS\footnote{Catalogue 
available under\\ {\it http://www.ipac.caltech.edu/2mass}}. These values are listed in the 
first row of Table \ref{bpl129}. The J-, H-, and K-band photometry was shifted from the
2MASS system to the CIT system using the colour transformations given by Carpenter 
(\cite{c01}). The apparent magnitudes were de-reddened adopting a reddening of $E_{B-V}=0.03$ 
(Stauffer et al. \cite{ssk98}) and the extinction law of Savage \& Mathis (\cite{sm79}). We 
adopt a distance of 133\,pc to the Pleiades, a value which is consistent with all recent 
measurements except for the Hipparcos value (see Pan et al. \cite{psk04}, Munari et al. 
\cite{mds04}). Thus, we obtain absolute magnitudes for BPL129 which are listed in Table 
\ref{bpl129}. 

The most probable age of the Pleiades is 125\,Myr (Stauffer et al. \cite{ssk98}). To estimate
the basic parameters of BPL129, we therefore compared the absolute magnitudes with the 125\,Myr 
isochrone of Baraffe et al. (\cite{bca98}). We obtain an object mass between 0.14 and $0.165\,M_{\odot}$, 
depending on the wavelength band in which the comparison is carried out. Averaging over all four 
available photometry points, gives us $0.156\,M_{\odot}$ as mass estimate of BPL129, corresponding 
to an effective temperature of $3205\pm100$\,K. Therefore, we use the model spectrum with 
$T_\mathrm{eff}=3200$\,K as SED for the photosphere of BPL129 (called $S_0$ in the following). 
Since the surface gravity for VLM stars in the Pleiades is best modelled with $\log{g}=5.0$ 
(Baraffe et al. \cite{bca98}), we used only the spectra for this gravity. 

The spots will have either lower or higher effective temperatures. We decided to vary the effective 
temperature of the spots $T_S$ in the simulation between 2000 and 4000\,K in steps of 100\,K. With 
this grid of temperatures, the contrast between photosphere and spots 
$(T_S - T_\mathrm{eff})/T_\mathrm{eff}$ varies in the range from 0-38\% for cool and 0-25\% for 
hot spots, which are typical values for solar-mass stars (Strassmeier \cite{s92}). The SED of 
the spots is called $S_S$ in the following. Since we do not have information about the spot 
distribution, we assume here that there is one spot on the surface of the target, whose temperature 
and size are to be determined. For the surface filling factor of the spot, i.e. the fraction of the 
surface of a hemisphere covered by the spot, we choose values between 0.01 and 0.5 in steps of 0.01. 

\begin{figure}[t]
\resizebox{9.0cm}{!}{\includegraphics[angle=-90,width=6.5cm]{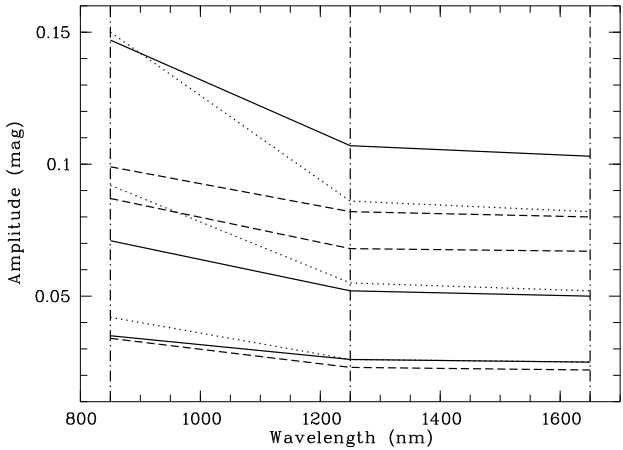}}
\caption{Photometric amplitudes in three wavelength bands as calculated in our simulations. 
Solid lines show the amplitudes for a constant spot temperature of $T_S=2700$\,K with filling 
factors 0.05, 0.1, 0.2 (from bottom to top). Dashed lines show the results for constant filling 
factor of 0.1, but for $T_S=2300$, 2500, and 3000\,K (from top to bottom). Dotted lines are for 
hot spots with constant filling factor of 0.1 and $T_S=3400$, 3600, 3800\,K (from bottom to top).}
\label{simul}
\end{figure}	

First we assume that the spot is cooler than the photosphere. Then in the light curve minimum, the 
spot will be on the front side, in the maximum on the backside. Under these conditions, the 
SED in the maximum $S_\mathrm{max}$ is just $S_0$ and in the minimum:
\begin{equation}
S_\mathrm{min} = (1.0 - f) \times S_0 + f \times S_S (T_S) 
\end{equation}
where $f$ is the filling factor of the spots and $T_S$ is between 2000 and 3200\,K. These 
SEDs for maximum and minimum are now convolved with transmission curves for the I-band and 
the CCD sensitivity curve, or with transmission curves for J- or H-band. By integrating over 
the respective SEDs, we compute the flux in the minimum and maximum in all three filters, 
called $F_\mathrm{min,I}$, $F_\mathrm{min,J}$, $F_\mathrm{min,H}$ and $F_\mathrm{max,I}$, 
$F_\mathrm{max,J}$, $F_\mathrm{max,H}$. The amplitude for the wavelength band X can 
then be calculated with:
\begin{equation}
A_X = -2.5 \log{\frac{F_\mathrm{min,X}}{F_\mathrm{max,X}}}
\end{equation}
These amplitudes were calculated for the grid of the parameters $f$ and $T_S$ given
above. A very similar calculation was carried out for hot spots. In this case, the SED in the
minimum is $S_0$, and the SED in the maximum is given by Equ. (1), now for $T_S>3200$\,K.
As the result of the simulations, we obtain a table giving $A_I$, $A_J$, and $A_H$
for the complete grid of spot temperatures and filling factors. 

In Fig. \ref{simul}, we show the amplitudes as a function of wavelength for a few selected 
combinations of $T_S$ and $f$. Solid lines give the amplitudes for a constant spot temperature
of $T_S=2700$\,K with filling factor 0.05, 0.1, 0.2. Dashed lines show the results for constant
filling factor of 0.1, but for $T_S=2300$, 2500, 3000\,K. Dotted lines are for hot spots with
constant filling factor of 0.1 and $T_S=3400$, 3600, 3800\,K. The plot clearly demonstrates 
that different spot parameters lead to significantly different photometric amplitudes. If either
the spot temperature or the filling factor is fixed, the amplitudes vary continuously with the
second parameter. On the other hand, the problem is degenerate in the sense that many combinations
of $T_S$ and $f$ lead to very similar amplitudes, which cannot be distinguished by observations.
Thus, we do not expect to be able to derive the exact values for $T_S$ and $f$ by comparing with
observations, but rather to constrain the parameter space. Hot spots show a much steeper decrease
of the amplitude towards red wavelengths, which can be used to distinguish them from cool spots. 

\begin{figure}[t]
\resizebox{9.0cm}{!}{\includegraphics[angle=-90,width=6.5cm]{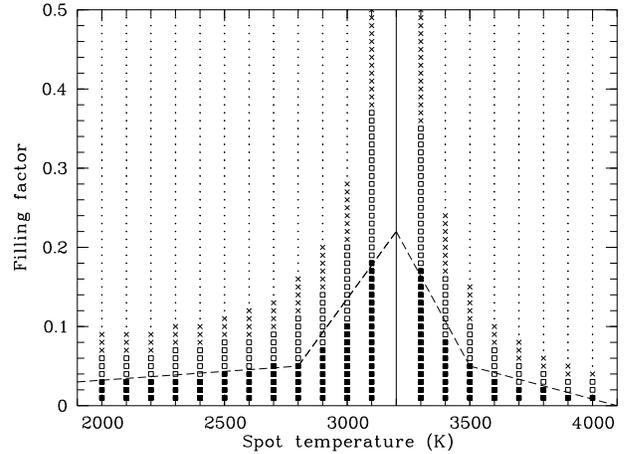}}
\caption{Constraints for the spot parameters from the observed I-band amplitudes: 
Filled squares show all possible combinations of filling factor and spot temperature
which would produce an I-band amplitude of $<0.035$\,mag, the upper limit of the
observed amplitudes. The dashed lines are a rough envelope of these values. Empty
squares are spot parameters for amplitudes between 0.035 and 0.07\,mag, crosses for
amplitudes between 0.07 and 0.1\,mag, small dots for values $>0.1$\,mag. The solid
line shows the photospheric temperature of our main target BPL129.}
\label{isim}
\end{figure}	

There are three limitations of our method: First, we can only investigate the
part of the spots, which is asymmetrically distributed, because the symmetric fraction of
the spot distribution does not contribute to the photometric variability. Thus, the
derived values for $T_S$ and $f$ give us only information about the 'asymmetric' spots.
This should not significantly bias the result for $T_S$, because there is no reason to
assume that the temperature of the spots on a particular object depends on their 
distribution. On the other hand, the values for the filling factor $f$ are only a lower
limit, since there could be much more spot coverage without any asymmetry.

The second problem is the unknown orientation of the rotational axis of the object. In the
simulation, we assume that the axis is perpendicular to the line of sight. An inclination
of the rotation axis would decrease the influence of the spots on the light curve, in the sense 
that the amplitude would decrease by a wavelength independent constant. Thus, the amplitudes 
which we derive in the simulation, are upper limits for the given spot parameters. The assumption
of an angle $i\approx 90$\degr\,between rotational axis and line of sight is, however, not 
unplausible for BPL129, which has the maximum I-band amplitude among our targets. Since the 
scatter of the amplitudes is partly caused by the scattering of $i$, we expect that objects with 
relatively high amplitudes (like BPL129) tend to have high inclination angles. 

Furthermore, it has been shown e.g. by Messina et al. (\cite{mrg01}) that light curve
amplitudes are related to the object rotation periods. Such a correlation
 -- if it also applies to our targets -- may introduce an additional scatter in the photometric 
amplitudes. A future determination of the inclination angles of the rotation axes of our targets 
should help to disentangle and quantify these effects, and requires additional observations.

\subsection{Comparison with I-band amplitudes}
\label{iband}

I-band amplitudes have been measured for nine VLM stars in the 
Pleiades (SE). All these values are below 0.035\,mag, which is the amplitude of BPL129, 
the star for which we derived amplitudes in the NIR. In addition, Terndrup et al. (\cite{tkp99}) 
present light curves for two VLM stars in this cluster: Whereas \object{HHJ-409} has an amplitude
of about 0.035\,mag, consistent with the upper limit in SE, the second star, \object{CFHT-PL-8},
has a higher amplitude of about 0.07\,mag. Since it is the only VLM object in the 
Pleiades with amplitude $>0.035$\,mag, whereas all other ten objects with photometric periods 
show lower amplitudes, we treat this amplitude of 0.07\,mag as an outlier, although this object 
clearly requires a more detailed investigation. Thus the upper limit for the photometric
amplitudes in the I-band is assumed to be 0.035\,mag. 

This can be used to constrain the parameter space for the spot properties. In Fig. \ref{isim}
we show spot temperature vs. filling factor for amplitudes $<0.035$\,mag (filled squares),
between 0.035 and 0.07\,mag (empty squares), between 0.07 and 0.1\,mag (crosses), and for
amplitudes $>0.1$\,mag (small dots). The solid line marks the photospheric temperature of 
BPL129. Our simulations were tailored for this object, in the sense that they are based
on its effective temperature, but we do not expect qualitatively different results for the
other objects: All known variable VLM objects in the Pleiades have effective temperatures
between 3400 and 2900\,K. Increasing or decreasing the values for $T_\mathrm{eff}$ by not
more than 300\,K would only shift the whole plot to higher or lower temperatures.

A dashed line marks the upper envelope of the filled squares and is an envelope for the range 
of spot parameters which could produce the amplitudes observed on VLM objects. Thus, the 
parameter space can be significantly constrained only with the I-band amplitudes: To explain 
the observed amplitudes, the filling factor has to be below 5\% (with a temperature difference
between 3 and 38\%) or the temperature difference has to be below 10\% (with a filling factor
between 1 and 19\%). If we include the 0.07\,mag amplitude of CFHT-PL-8 (see above), even 
the solutions shown in open squares have to be considered, enlarging the spectrum of possible 
solutions. In this case, the filling factor has to be below 10\% (with a temperature difference
between 3 and 38\%) and/or the temperature contrast has to be below 22\% (with filling factors 
between 1 and 37\%).

\subsection{Constraints from NIR time series}
\label{nirband}

Now we compare the results of our multi-filter results of BPL129 with
the simulations. The amplitudes in I, J, and H are 0.035, 0.035, and 0.032\,mag. 
A first evaluation of these values is possible with Fig. \ref{simul}.
Since the photometric amplitudes are below 0.04\,mag, they can only be explained 
with low contrast between spots and photosphere, but filling factors up to 10\%, or 
larger temperature contrast and filling factors below 5\%, as already shown in 
Sect. \ref{iband}. On the other hand, the amplitude does not show a sharp
decrease from the I- to the J-band, indicating that the spots are probably cooler
than the photosphere.

\begin{figure}[t]
\resizebox{9.0cm}{!}{\includegraphics[angle=-90,width=6.5cm]{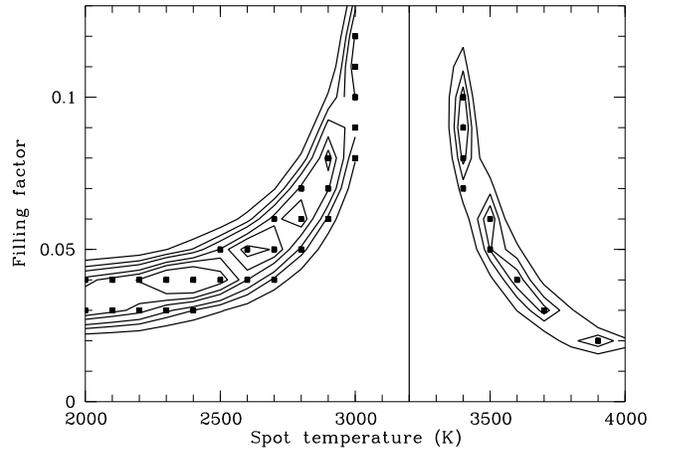}}
\caption{Contour plot for the $\chi^2$ values from the comparison between observed
and simulated amplitudes. Contour lines start at $\chi^2=$3.0 and are plotted for
$\chi^2=$3.0, 2.0, 1.5, 1.0, 0.75, 0.5, indicating increasing quality of the fit. 
Filled squares show all combinations of spot temperature and filling 
factor which would produce amplitudes within the error bars of our observations.
The vertical solid line indicates the photospheric temperature of BPL129. Note that the
hot spot solutions on the right side never reach to the 4th contour line, and $\chi^2$ is
larger than 0.9 everywhere. They are thus significantly worse than the cool spot solutions on 
the left side. (One datapoint with $\chi^2 = 0.92$ at $T_S = 3300$\,K and $f = 19$\% is 
not shown in the figure.)}
\label{contour}
\end{figure}   

In Fig. \ref{contour} we show with filled squares all combinations of spot temperature and 
filling factor from our simulations, which are within the error bars of the measured amplitudes. 
Since our error estimates are very conservative, we can safely exclude that the spot
parameters of BPL129 are not represented by this set of parameters.
In agreement with the result from Sect. \ref{iband} we find that either the filling factor 
has to be between 2 and 5\% or the temperature difference between spots and photosphere has
to be lower than 10\% to explain the observed amplitudes. 

To assess the statistical significance of these results, we calculated $\chi^2$ for
all data points: (obs - observed amplitudes, sim - simulated amplitudes)
\begin{equation}
\chi^2 = \sum_{X=I,J,H} \Bigl| \frac{A_\mathrm{X,obs} - A_\mathrm{X,sim}}{\sigma_\mathrm{X}} \Bigr|^2
\end{equation}
In Fig. \ref{contour} we show a contour plot of the results, with contours for $\chi^2=$
3.0, 2.0, 1.5, 1.0, 0.75, 0.5, indicating increasing quality of the fit. The five solutions which 
deliver the minimum $\chi^2$ ($<0.5$) and 
thus the best agreement between simulated and observed amplitudes have filling factors of 4-5\% and 
a spot temperature between 2200 and 2600\,K, corresponding to a temperature contrast of 18-31\%. 
The most probable spot configuration has $T_S = 2400$\,K and $f = 4$\%. Note that hot spot
solutions show only up to three contour lines and give $\chi^2$ values $>0.9$. They are thus 
clearly less signficant than the best solutions with cool spots.

Recapitulating, we find that the best agreement between simulated and observed amplitudes
is reached with very low filling factors of 4-5\%. We like to stress once more that this is 
not the total surface filling factor. This value is only related to the asymmetrically distributed
fraction of the spots. A low value for the filling factor means that either the total 
spot coverage is low or the spot distribution is rather symmetric. The best solutions are
found for cool spots, where the temperature contrast between spots and photosphere 
is probably 18-31\%. 

\subsection{Comparison with literature data}
\label{comp}

Our results are now compared to literature data for more massive stars. Spot parameters 
have been determined for a large number of solar-mass stars, at various evolutionary stages 
(see Bouvier \& Bertout \cite{bb89}, Strassmeier \cite{s92}). Both pre-main sequence and 
main sequence stars show a very large range of filling factors and spot temperatures. The 
temperature contrast varies between 5 and 40\%, where the majority of the values is between 
20 and 30\%. Thus, the result of 18-31\% which we obtain for BPL129 is not untypical when 
compared to solar-mass stars. On the other hand, according to Strassmeier \cite{s92}, 
solar-mass stars show filling factors between 0 and 25\% of the entire stellar surface. 
In our simulations, the filling factor is defined as the fraction of the stellar 
hemisphere covered with spots (see Sect. \ref{desc}). Hence, the value derived for 
BPL129 (4-5\%) would correspond to 2-3\% if related to the entire stellar surface, and
is thus at the low end of the range found for solar-mass stars. We come to the conclusion 
that BPL129 shows rather low filling factors compared to more massive stars.

As discussed in Sect. \ref{intro}, there are at least two possible explanations for these results.
The low filling factor measured for BPL129 might be caused by a change of the interior dynamo. 
It has often been discussed that the solar-type $\alpha\Omega$ dynamo, which operates in a shell 
between convective and radiative zone, might be replaced by a distributed dynamo in fully convective
objects. This dynamo could be either of turbulent nature (Durney et al. \cite{ddr93}) or 
based on Coriolis forces, the so-called $\alpha^2$ dynamo (Kueker \& Ruediger \cite{kr99}).
A distributed dynamo should lead to a more symmetric spot distribution, which would decrease
the filling factor of the asymmetric component of the spots (see Sect. \ref{nirband}). 

Decreasing effective temperatures could also cause a low filling factor, 
because the cooler the objects, the less efficient is the coupling between magnetic
field and gas. Indeed, a decline of the chromospheric and coronal activity is observed at late M
spectral types (Mohanty \& Basri \cite{mb03}, Mokler \& Stelzer \cite{ms02}). If the same decline
does also occur in the photospheric regime, we should expect very few spots on objects with
spectral type later than M9. It is, however, unlikely that the difference in effective 
temperature is the reason for the low filling factor measured for BPL129, since
BPL129 is of mid-M spectral type, a region where VLM objects still show high levels of
chromospheric and coronal activity. Thus, the low filling factor of BPL129 is more probably
the consequence of a change of the interior dynamo than of the low effective temperature.

As already mentioned in Sect. \ref{intro}, Terndrup et al. (\cite{tkp99}) estimated spot parameters
for HHJ-409, a Pleiades VLM star with $M=0.39\,M_{\odot}$. They find a filling factor of 13\% 
and a temperature contrast of 6\%. These spot properties are clearly different from 
the best solutions we have found for BPL129 (see Sect. \ref{nirband}), and are in addition
not contained among the solutions which are in agreement with the errorbars of our 
photometry (filled squares in Fig. \ref{contour}). Thus, based on the multi-filter photometry 
HHJ-409 has spot parameters different from those of BPL129.

One explanation may be that Terndrup et al. (\cite{tkp99}) used a simple blackbody distribution 
to model the spectrum of photosphere and spots, which might be an unrealistic approach, given 
that the SEDs of very cool objects deviate strongly from a blackbody. On the other hand, the 
mass difference between BPL129 and HHJ-409 is significant. In particular, the mass of HHJ-409 
is probably slightly above the limit where the objects become fully convective (see Sect. 
\ref{intro}) leading to a different type of dynamo mechanism in operation. As explained above,
this might be the reason for the different spot characteristics.

\section{Conclusions}
\label{conc}

We investigate the spot properties of VLM objects by comparing the amplitudes of photometric
variations in three wavelength bands with simulations. This study is based on the variability
study in the Pleiades, which delivered I-band amplitudes for nine VLM stars (SE). All these
amplitudes are $\le 0.035$\,mag. These VLM stars have been additionally monitored in the 
J- and H-band. For the star BPL129, which has an I-band amplitude of 0.035\,mag, we were able 
to derive amplitudes of 0.035 and 0.032\,mag in the J- and H-band. 

The simulations are based on model spectra from Allard et al. (\cite{ahs00}), which give
us the SED for the photosphere and for various spot temperatures. We compute
the amplitude between the light curve minimum and maximum depending on the spot temperature
and the filling factor by assuming that the star has one spot co-rotating with the object and
the rotational axis is perpendicular to the line of sight. The filling factor in these
simulations is determined by the fraction of the stellar hemisphere which is covered 
asymmetrically with spots. 

In a first step, we compare only the I-band amplitudes with these models. We find that the
filling factor has to be below 5\% and/or the temperature contrast has to be below 10\%
to produce the observed I-band amplitudes. The data points in the J- and H-band for BPL129
can be used to further constrain the spot properties. We find best agreement between
observed and simulated amplitudes with cool spots with temperatures 18-31\% below the
photospheric temperature and a filling factor of 4-5\%. 

These results are compared to similar studies for more massive stars.
It turned out that spots on VLM stars have similar temperature contrast, but rather low
filling factors when compared to solar-mass stars. This might be the consequence of a change
of the dynamo from a solar-type shell dynamo to a distributed dynamo in these fully
convective objects.

\begin{acknowledgements}
      We are grateful to Jens Woitas and Nicolas Cardiel, who actively supported the MAGIC 
      observations on Calar Alto. It is a pleasure to acknowledge the help of Artie Hatzes and 
      Rafael Rebolo, who supported the organisation of the simultaneous observations 
      for this project. We thank the referee, S. Messina, for his helpful comments.
      Part of this work was supported by the German
      \emph{Deut\-sche For\-schungs\-ge\-mein\-schaft, DFG\/} project
      numbers Ei~409/11--1, Ei~409/11--2. A.S. received travel funds from
      the DFG HA3279/2--1 project. 
      D.F. received financial support from the Cosmo-Grid project, funded
      by the Program for Research in Third Level Institutions under the National
      Development Plan and with assistance from the European Regional Development
      Fund. The publication makes use of data products from the Two Micron All 
      Sky Survey, which is a joint project of the University of Massachusetts and 
      the Infrared Processing and Analysis Center/California Institute of Technology, 
      funded by the National Aeronautics and Space Administration and the National 
      Science Foundation.
\end{acknowledgements}


\begin{thebibliography}{}

\bibitem[1997]{aha97} Allard, F., Hauschildt, P. H., Alexander, D. R., Starrfield, S.,
1997, ARA\&A, 35, 137

\bibitem[2000]{ahs00} Allard, F., Hauschildt, P. H., Schwenke, D., 2000, ApJ, 540, 1005

\bibitem[2001]{aha01} Allard, F., Hauschildt, P. H., Alexander, D. R., Tamanai, A., 
Schweitzer, A., 2001, ApJ, 556, 357

\bibitem[1998]{bca98} Baraffe, I., Chabrier, G., Allard, F., Hauschildt, P. H., 1998, 
A\&A, 337, 403

\bibitem[2001]{bc01} Barnes, J. R., Collier Cameron, A., 2001, MNRAS, 326, 950

\bibitem[1989]{bb89} Bouvier, J., Bertout, C., 1989, A\&A, 211, 99

\bibitem[1995]{bck95} Bouvier, J., Covino, E., Kovo, O., Martin, E. L., Matthews, J. M., 
Terranegra, L., Beck, S. C., 1995, A\&A, 299, 89

\bibitem[2001]{c01} Carpenter, J. M., 2001, AJ, 121, 2851

\bibitem[1997]{cb97} Chabrier, G., Baraffe, I., 1997, A\&A, 327, 1039

\bibitem[1993]{ddr93} Durney, B. S., De Young, D. S., Roxburgh, I. W., 1993, SoPh, 145, 207

\bibitem[1981]{fd81} Feigelson, E. D., Decampli, W. M., 1981, ApJ, 243, 89

\bibitem[1996]{hsd96} Henry, T. J., Soderblom, D. R., Donahue, R. A., Baliunas, S. L.,
1996, AJ, 111, 439

\bibitem[1993]{hbb93} Herbst, T.M., Beckwith, S.V., Birk, C., Hippler, S., 
McCaughrean, M.J., Mannucci, F. \& Wolf, J. 1993, SPIE, 1946, 605

\bibitem[1994]{hhg94} Herbst, W., Herbst, D. K., Grossman, E. J., Weinstein, D., 1994,
AJ, 108, 1906

\bibitem[2001]{hbm01} Herbst, W., Bailer-Jones, C. A. L., Mundt, R., ApJ, 554, 197

\bibitem[2002]{hbm02} Herbst, W., Bailer-Jones, C. A. L., Mundt, R., Meisenheimer, K., Wackermann, R.,
2002, A\&A, 396, 513

\bibitem[1999]{kr99} K{\"u}ker, M., R{\"u}diger, G., 1999, A\&A, 346, 922

\bibitem[2003]{l03} Lamm, M. H., 2003, Ph.D. thesis, University of Heidelberg

\bibitem[2004]{lbm04} Lamm, M. H., Bailer-Jones, C. A. L., Mundt, R., Herbst, W., Scholz, A.,
2004, A\&A, 417, 557

\bibitem[2005]{lmb05} Lamm, M. H., Mundt, R., Bailer-Jones, C. A. L., Herbst, W., 
2005, A\&A, 430, 1005

\bibitem[2001]{mrg01} Messina, S., Rodono, M., Guinan, E. F., 2001, A\&A, 366, 215

\bibitem[2003]{mb03} Mohanty, S., Basri, G., 2003, ApJ, 583, 451

\bibitem[2002]{ms02} Mokler, F., Stelzer, B., 2002, A\&A, 391, 1025

\bibitem[2004]{mds04} Munari, U., Dallaporta, S., Siviero, A., Soubiran, C., Fiorucci, M., 
Girard, P., 2004, A\&A, 418, 31

\bibitem[2004]{psk04} Pan, X., Shao, M., Kulkarni, S. R., 2004, Nature, 427, 326

\bibitem[1975]{p75} Parker, E. N., 1975, ApJ, 198, 205

\bibitem[2000]{phj00} Pinfield, D. J., Hodgkin, S. T., Jameson, R. F., Cossburn, M. R., Hambly, N. C., 
Devereux, N., 2000, MNRAS, 313, 347

\bibitem[1990]{ptv90} Piskunov, N. E., Tuominen, I., Vilhu, O., 1990, A\&A, 230, 363

\bibitem[1979]{sm79} Savage B.H., Mathis J.S., 1979, ARA\&A, 17, 73

\bibitem[2004]{s04} Scholz, 2004, Ph.D. thesis, University of Jena

\bibitem[2004a]{se04a} Scholz, A., Eisl{\"o}ffel, J., 2004, A\&A, 419, 249 

\bibitem[2004b]{se04b} Scholz, A., Eisl{\"o}ffel, J., 2004, A\&A, 421, 259 (SE)

\bibitem[2005]{se05} Scholz, A., Eisl{\"o}ffel, J., 2005, A\&A, 429, 1007

\bibitem[1984]{smw84} Schrijver, C. J., Mewe, R., Walter, F. M., 1984, A\&A, 138, 258

\bibitem[1985]{s85} Soderblom, D. R., 1985, AJ, 90, 2103

\bibitem[1980]{sw80} Spiegel, E. A., Weiss, N. O., 1980, Nature, 287, 616

\bibitem[1995]{sed95} Stanford, S.A., Eisenhardt, P.R.M, Dickinson, M. 1995, ApJ, 450, 512

\bibitem[1994]{scg94} Stauffer, J. R., Caillault, J.-P., Gagne, M., Prosser, C. F., Hartmann, L. W.,
1994, ApJS, 91, 625

\bibitem[1998]{ssk98} Stauffer, J. R., Schultz, G., Kirkpatrick, J. D., 1998, ApJ, 499, 199

\bibitem[2000]{snh00} Stelzer, B., Neuh{\"a}user, R., Hambaryan, V., 2000, A\&A, 356, 949

\bibitem[1992]{s92} Strassmeier, K. G., 1992, ASP Conf. Ser., 34, 39

\bibitem[1999]{tkp99} Terndrup D. M., Krishnamurthi A., Pinsonneault M. H., Stauffer J. R., 1999, 
AJ, 118, 1814

\bibitem[1987]{vph87} Vogt, S. S., Penrod, G. D., Hatzes, A. P., 1987, ApJ, 321, 496

\bibitem[1978]{w78} Wilson, O. C., 1978, ApJ, 226, 379

\end{thebibliography}
\end{document}